\documentstyle[11pt,a4,ere99]{article}
\input{epsf}
\newcommand{\gsim}{\stackrel{>}{\sim}}
\newcommand{\lsim}{\stackrel{<}{\sim}}

\begin{document}

\title{Cosmic Microwave Background Anisotropies:\\
       Inflation versus Topological Defects\\
\emph{Proceedings of the Spanish Relativity Meeting}\\ Bilbao, 1999}

\author{Mairi Sakellariadou}

\address{
Institut des Hautes Etudes Scientifiques, 91440 Bures-sur-Yvette, France\\
DARC, Observatoire de Paris,
UPR 176 CNRS, 92195 Meudon Cedex, France\\
E-mail: {\em {\tt mairi@amorgos.unige.ch}}}

\maketitle\abstract{I present a briefly summary of the current
status of inflationary models versus topological defects
scenarios, as the mechanisms which could have induced the initial density
perturbations, which left an imprint on the cosmic microwave
backgound radiation anisotropies.}

\section{Introduction}

The origin of the large scale structure in the universe, remains 
one of the most important questions in cosmology.  Within the framework
of gravitational instability, there are two currently investigated
families of models to explain the formation of the observed structure.  
Initial density perturbations can either be due to ``freezing in'' of 
quantum fluctuations of a scalar field during an inflationary 
period~\cite{stein}, or they may be seeded by a class of topological defects, 
which could have formed naturally during a symmetry breaking phase transition 
in the early universe~\cite{kibble}.
The cosmic microwave background radiation (CMBR) anisotropies provide a link 
between theoretical predictions and observational data, which may allow us
to distinguish between inflationary models and topological defects scenarios, 
by purely linear analysis.   

The CMBR, last scattered at the epoch
of decoupling, has to a high accuracy a black-body distribution~\cite{bb}, 
with a temperature $T_0=2.728\pm 0.002 ~{\rm K}$, which is
almost independent of direction.  The DMR experiment on the COBE
satellite measured a tiny variation in intensity of the CMBR, at fixed
frequency. This is equivalently expressed as a variation ${\rm \delta}T$ 
in the temperature, which was measured~\cite{cobe} to be 
${\rm \delta }T/T_0 \approx 10^{-5}$.  
The 4-year COBE data are fitted by a scale-free
spectrum; the spectral index was found to be $n_{\rm S}=1.2\pm 0.3$
and the quadrupole anisotropy~\cite{cobe}
 $Q_{\rm rms-PS}=15.3^{+3.8}_{-2.8} ~{\rm\mu K}$.

I divide my talk into three parts, namely sections 2, 3 and 4. 
In Section 2, I define the angular power spectrum in terms of 
$\delta T/T$ and describe the two families of models which give $\delta T/T$. 
I then briefly discuss why we need such ``exotic'' models to address the 
issue of large scale structure formation and I compare topological defects 
models versus inflationary ones.
In Section 3, my discussion becomes a bit more technical. I describe the 
physical mechanisms which perturb the CMBR on different angular scales
and present the predictions of each family of models for large/small
angular scales. In Section 4, I discuss the ``rigidity''of the two 
families of models and present results from specific cases.
I then list the lessons we have already learned from the  CMBR anisotropies 
measurements and what we expect to learn from future experiments.
I close with the conclusions given in Section 5.

\section{The plot: presenting the problem}
\subsection{The angular power spectrum in terms of $\delta T/T$}
\label{subsec:prod}

We want to calculate temperature anisotropies in the sky, thus it is natural
to expand $\delta T/T$ in spherical harmonics:
\begin{equation}
{\delta T\over T}({\bf n}) =
\sum_{\ell m} a_{\ell m} Y_{\ell m}({\bf n})~.
\end{equation}
The angular power spectrum of CMBR anisotropies is expressed in terms
of the dimensionless coefficients $C_\ell$, which appear in the
expansion of the angular  correlation function in terms of the
Legendre polynomials $P_\ell$:
\begin{equation}
\langle{\delta T\over T}({\bf n}){\delta T\over T}({\bf n}')
\rangle\left|_{{~}_{\!\!({\bf n\cdot n}'=\cos\vartheta)}}\right. =
  {1\over 4\pi}\sum_\ell(2\ell+1)C_\ell P_\ell(\cos\vartheta)~.
\end{equation}
It compares points in the sky separated by an angle $\vartheta$.
Here the brackets denote spatial average, or expectation values if
perturbations are quantized.
The value of $C_\ell$ is determined by fluctuations on angular scales
of order $\pi/\ell$. The angular power spectrum of anisotropies
observed today is usually given by the power per logarithmic interval
in $\ell$, plotting $\ell(\ell+1)C_\ell$ versus $\ell$.
The coefficients $C_\ell$ are related to $a_{\ell m}$ by
\begin{equation}
C_\ell = {\langle\sum_m | a_{\ell m} |^2 \rangle \over 2\ell+1}~.
\end{equation}

\subsection{The two families of models which give $\delta T/T$}

The inflationary paradigm was proposed in
order to explain the shortcomings of the standard (Big Bang)
cosmological model. In addition, it offers a scenario for the
generation of the primordial density perturbations, which can lead to
the formation of the observed large scale structure.
Within the inflationary paradigm, a possible mechanism 
for the generation of the angular power spectrum of CMBR anisotropies and the 
creation of the large scale structure is based on the quantum fluctuations 
that exited the horizon during inflation~\cite{freeze}.
These fluctuations are the source of the primordial spectrum of
density inhomogeneities~\cite{denpert}, which has left an imprint on the CMBR. 
The observed large scale structure could have been generated by the growth 
through gravitational instability of this  primordial spectrum of 
perturbations in the otherwise uniform  distribution of matter.

Alternatively,  CMBR anisotropies could be triggered by topological defects,
which could have been formed naturally during a symmetry breaking phase 
transition in the early universe. Any global defects or local cosmic strings
could, in principle, induce the initial perturbations. The measurements 
of the COBE satellite 
provide the normalization $T_{\rm c}^2/M_{\rm Pl} \sim 10^{-5}$,
where $T_c$ denotes the temperature at the phase transition. Thus,
to seed the observed large scale structure, topological 
defects should have been formed at $ T_{\rm c}\sim 10^{16}$ GeV. 

\subsection{Why do we need such ``exotic'' models?}

I will briefly sketch why structure formation in a baryonic hot Big Bang
model is incompatible with the smoothness of the CMBR. As you know, expansion 
counteracts gravitational attraction, and therefore gravitational 
instabilities 
do not grow exponentially. In a radiation-dominated universe, radiation 
pressure inhibits any significant growth of $\delta\rho/\rho$. After the
transition to the matter-dominated era, $\delta\rho/\rho$ grows like the scale
factor, implying $\delta\rho/\rho\lsim a_0/a_{\rm eq}\sim 10^3 - 10^4$.
Thus, initial fluctuations of order $\sim 10^{-3} - 10^{-4}$ are required
at the time of the radiation- to the matter-dominated era transition, 
in the matter component, which enhanced by gravity
led to $\delta\rho/\rho\sim 1$ and, thus, to the observed large scale 
structure. Since such
fluctuations are incompatible with the observed smoothness of the CMBR, we
deduce the need for non-baryonic matter.

\subsection{Topological defects models versus inflationary ones}

Inflationary fluctuations are produced at a very 
early stage of the evolution of the universe, and are driven far beyond the 
Hubble radius by inflationary expansion. Subsequently, they are not
altered anymore and evolve freely according to homogeneous linear
perturbation equations until late times. These fluctuations are termed
``passive'' and ``coherent''~\cite{joao}. ``Passive'', since no new
perturbations are created after inflation; ``coherent'' since randomness
only enters the creation of perturbations during inflation,
subsequently they evolve in a deterministic and coherent manner.
The linear evolution of these acausal (there exist correlations on super-Hubble
scales) and coherent initial perturbations has been explored in a large 
number of inflationary models.

On the other hand, in models with topological defects or other types
of seeds, fluctuations are  generated continuously and evolve
according to inhomogeneous linear perturbation equations. The seeds 
are defined as
 any   non-uniformly distributed form of energy, which 
contributes only a small fraction to the total energy density of the universe
and which interacts with the cosmic fluid only gravitationally.
The energy momentum tensor of the seeds  is determined by the topological 
defects (seeds) evolution which, in general,
is a non-linear process. These perturbations are called ``active'' 
and ``incoherent''~\cite{joao}. ``Active'' since new fluid perturbations are
induced continuously due to the presence of the seeds;
``incoherent'' since the randomness of the
non-linear seed evolution which sources the perturbations can destroy
the coherence of  fluctuations in the cosmic fluid.
The highly non-linear structure of the topological defects dynamics makes the
study of the evolution of these causal (there are no correlations on
super-Horizon scales) and incoherent initial perturbations much more 
complicated. That is why the number of models with seeds in which CMBR 
anisotropies have been addressed is rather limited. 

While the 
predictions for CMBR anisotropies in the context of inflationary models are 
quite robust, the predictions within topological defects models depend on 
the details of the particular numerical simulations.
In addition, even though one expects that only the scalar mode will
produce more power on small angular scales than on larger ones,
in topological defects models all three modes --- scalar,
vector and tensor ones --- contribute to the overall CMBR anisotropies.

\section{The plot thickens: the employed methodology}

\subsection{The physical mechanisms which perturb the 
 cosmic microwave background on different angular scales}

If we neglect  Silk damping in a first step and integrate the photon
geodesics in the perturbed metric,
gauge invariant linear perturbation analysis leads to~\cite{d90,RuthReview}   
\begin{equation}
{\delta T\over T}({\bf x}, {\bf n}) = \left[
  {1\over 4}D_g({\bf x}) +V_j({\bf x})n^j 
	+(\Psi-\Phi)({\bf x})\right]_{\eta_{\rm dec}}  
	+ \int_{\eta_{\rm dec}}^{\eta_{\rm now}} (\dot{\Psi} - \dot{\Phi} )
({\bf x}, \eta) d \eta~, 
\label{dT} 
\end{equation}
where over-dot denotes derivative with respect to conformal time $\eta$.
$\Phi$ and $\Psi$ are the Bardeen potentials, quantities describing the 
perturbations in the geometry, $\bf V$ is the peculiar velocity of 
the baryon fluid with respect to the overall Friedman expansion and
$D_g$ specifies the intrinsic density fluctuation in the radiation 
fluid.  

The first term in Eq.~(\ref{dT}) describes the intrinsic inhomogeneities
on the surface of the last scattering due to acoustic oscillations
prior to decoupling. It also contains  contributions 
to the geometrical perturbations. The second term describes 
the relative motions of emitter and 
observer. This is the Doppler contribution to the 
CMBR anisotropies. It appears on the same angular scale as the
acoustic term and we  denote the sum of the acoustic
and Doppler contributions by  ``acoustic peaks''. The 
last two terms are due to the inhomogeneities in the space-time 
geometry; the first contribution determines the change in the photon
energy due to the difference of the gravitational
potential at the position of emitter and observer. Together with the part
contained in $D_g$ they represent the ``ordinary'' Sachs-Wolfe 
effect. The second term accounts for red-shifting or blue-shifting
caused by the time dependence of the gravitational field along the 
path of the photon (Integrated Sachs-Wolfe (ISW) effect). The sum of
the two terms is the full  Sachs-Wolfe contribution (SW).

On angular scales  
$0.1^\circ\stackrel{<}{\sim}  \theta\stackrel{<}{\sim}  2^\circ$, 
the main contribution to the CMBR anisotropies comes from the acoustic peaks,
while the SW effect is  dominant  on large angular
scales. For topological defects models,  the gravitational contribution is
mainly due to the ISW. The ``ordinary'' Sachs-Wolfe term even has 
the wrong spectrum, a white noise spectrum instead of a
Harrison-Zel'dovich~\cite{ruthzhou} spectrum.

On scales smaller than about $0.1^o$,  the anisotropies
are damped due to the finite thickness of the recombination shell,
as well as by photon diffusion during recombination (Silk damping).
Baryons and photons are very tightly coupled before recombination and
oscillate as one component fluid.
During the process of decoupling, photons slowly diffuse out of
over-dense into under-dense regions. To fully account for this
process, one has to solve the Boltzmann equation.

\subsection{Predictions of each family of models for large/small
angular scales}

The simplest topological defects models of structure
formation show conflicts with observational data. As it was first
shown in Ref.~[11], global topological defects models predict
strongly suppressed acoustic peaks. While on large angular scales the
predicted CMBR spectrum is in
good agreement with COBE measurements, on smaller angular scales the
topological defects models cannot reproduce the data of the Saskatoon
experiment, namely the characteristics of the first acoustic peak.  
One can manufacture models~\cite{ruthmairi} with
structure formation being induced by {\sl scaling} seeds, which lead to an
angular power spectrum with the same characteristics (position and
amplitude of acoustic peaks), as the one predicted by standard
inflationary models.
(By scaling seeds, we mean that in the absence of any intrinsic length 
scale other than the cosmic horizon, the behavior of the seed functions 
in terms of which we parametrize the energy momentum tensor of the seeds, is 
determined by dimensional reasons~\cite{ruthmairi}. The seed functions of
scaling sources have white noise spectra~\cite{ruthmairi}.)
 The open question is, though, whether such models
are the outcome of a realistic theory.  At this point, I would like to 
remind you that the question whether or not inflationary models which fit
the data are 
physical or not, has also to be addressed, even though people often tend to 
forget about it.

 On large angular scales, both families of models predict an approximately
scale-invariant Harrison-Zel'dovich spectrum~\cite{h,z}, with however a 
different
prediction regarding the statistics of the 
induced perturbations. Naturally, topological defects generate non-Gaussian 
perturbations, whereas  non-Gaussian fluctuations go beyond the paradigm of 
cold dark matter (CDM) and slow roll inflation. 
Recently, the COBE data have been used to test the gaussianity of
the CMBR anisotropies. Three groups~\cite{non-gaus1,non-gaus2,non-gaus3} 
have  reported results showing that the fluctuations would not be Gaussian. 
From a theoretical point of view, once non-vacuum initial states for 
cosmological 
perturbations are allowed~\cite{jam}, a generic prediction~\cite{jam} is indeed
deviations from gaussianity in the CMBR map.

I will schematically show~\cite{rd} how both families of models 
predict a scale-invariant
Harrison-Zel'dovich spectrum on large angular scales, 
while their predictions regarding the characteristics of the
first acoustic peak differ a
lot. Let us start with the large angular scales. For inflationary models,
$\Phi = -\Psi$ and $\dot{\Psi} = 0$. In Fourier space, 
$\langle|\Psi(k)|^2\rangle\propto1/k^3$. On super-horizon ($k\eta\ll 1$) 
scales,
\begin{equation}
{\delta T\over T}({\bf n,k}) = {1\over 3}\exp(i{\bf n\cdot k}\eta_{now})
\Psi(\eta_{dec},{\bf k})~.
\end{equation}
Using 
$$C_\ell = {2\over \pi}\int dk k^2{\langle|\Delta_{\ell}(k)|^2
\rangle\over(2\ell+1)^2~}
\mbox{~~~~where~~~~}
\Delta_\ell(k)={2\ell +1\over 3}\Psi(k,\eta_{dec})j_\ell(k\eta_{now})~,
$$
one finds that on large angular scales $C_\ell \propto [\ell(\ell+1)]^{-1}$.

Let us now turn to the case of topological defects models. In Fourier space,
the Bardeen potentials $\Phi, \Psi$ have white noise spectra on super-horizon
scales. On $k\eta\ll 1$, we get $\langle|\Psi(k)|^2\rangle \propto \eta^3$.
Since $D_g \sim (k\eta)^2\Psi$, $D_g$ is negligible on 
$k\eta\ll 1$.
At $\eta\sim 1/k$, the defects enter the horizon and their contribution 
decays. I use 
$$
C_\ell = {2\over \pi}\int dk k^2{\langle|\Delta_{\ell}(k)|^2
\rangle\over(2\ell+1)^2}~,\nonumber
$$
$$
\mbox {where} ~~~ {1\over 2\ell+1}\Delta_\ell(k)=
2\Psi(\eta_{dec},k)j_\ell(k\eta_{now}) + 2\int_{\eta_{dec}}^{\eta_{now}}
	\dot{\Psi}(t,\eta)j_\ell(k(\eta_{now}-\eta))d\eta~,
$$
and consider that the dark matter of the universe is CDM. 
Once the perturbation enters the horizon, it is dominated by the 
contribution due to CDM, which is time-independent. Since $\dot{\Psi}$ 
vanishes for $k\eta>1$, I perform the integration only until $\eta=1/k$.
Neglecting the weak time dependence of $j_\ell$ in the interval of 
integration, one can easily check that
$$
\Delta_\ell(k)=2(2\ell+1)\Psi(\eta=1/k,k)j_\ell(k\eta_{now})~,
~~
\mbox{with} ~~~~~\langle|\Psi(\eta=1/k,k)|^2\rangle\propto k^{-3}~.
$$
The ISW term behaves like the inflationary SW contribution
leading to
$C_\ell \propto [\ell(\ell+1)]^{-1}$ on large angular scales.
Thus, both families of models predict scale-invariant  
Harrison-Zel'dovich  spectra on large angular scales.

Let us now study~\cite{rd} the intermediate angular scales, where the 
first acoustic peak
appears. Energy and momentum conservation leads to
\begin{equation}
\dot{D}_g=- {4\over 3}kV~~,~~~~ \dot{V}=2k\Psi
	+{1\over 4}kD_g~,
\end{equation}
\begin{equation}
\mbox{implying}~~~~~~~
\ddot{D}_g +{1\over 3}k^2D_g=-{8\over 3}k^2\Psi~.
\end{equation}
In the case of topological defects models,
since
\begin{equation} 
\Psi \propto \left\{\begin{array}{lll}
\eta^{3/2} &\mbox{~~on }&k\eta\ll 1\\
k^{-3/2} &\mbox{~~on }&
k\eta\gg 1
\end{array} \right.
\end{equation}
one obtains
\begin{equation}
D_g\sim \left\{\begin{array}{lll}
(k\eta)\Psi &\mbox{~~ on }&k\eta\ll 1\\
\Psi(\cos{k\eta\over\sqrt{3}}-1)&\mbox{~~ on }&
k\eta \gg 1
\end{array} \right.
\mbox{~for topological defects models.}
\end{equation}
Thus, $D_g$ is  very small at horizon crossing
and first has to grow to achieve its maximum.

In the case of inflationary models,
\begin{equation}
\ddot{D}_g +{1\over 3}k^2D_g=-{8\over 3}k^2\Psi~.
\end{equation}
Since $\Psi$ is constant (time independent) on $k\eta \ll 1$, one obtains
\begin{equation}
D_g\sim \left\{\begin{array}{lll}
    \Psi &\mbox{~~ on }&k\eta\ll 1\\
    -\Psi (\sin{k\eta\over\sqrt{3}}-1)&\mbox{~~ on }&
k\eta\gg 1
\end{array} \right.
\mbox{~for adiabatic inflationary models.}
\end{equation}
$D_g$ is at its  maximum on $k\eta\ll 1$ and 
starts decaying at horizon crossing.
That is why, the first acoustic peak is displaced to larger 
$\ell$ (smaller angular scales) for topological defects models.

Summarizing, we can say that the difference between the two
classes of theories, is that while the ISW is mostly negligible
in inflationary models, it is present in topological defects scenarios,
leading to an increase in the power spectrum on large angular scales.

\section{The plot unfolds: results and discussion}

\subsection{The ``rigidity''of the two families of models}

One my ask to which degree the predictions of the two competing classes of 
theories 
are indeed rigid. Of course, one can always add epicycles which will lead to a 
different behavior.
For example, within the context of an inflationary scenario, one can push the 
first acoustic peak to larger $\ell$'s, considering an open 
universe. Since to achieve that, one has to introduce two scalar fields, 
inflation, to my mind, looses some of its elegance (simplicity). 
To discriminate 
between topological defects models with $\Omega=1$ and open inflationary 
models, 
one has to find the position $\ell_2$ of the second acoustic peak.
\begin{equation}
\Delta\ell\equiv \ell_2-\ell_1=[\Delta\ell(\Omega=1)] ~\Omega^{-1/2}\sim
300 ~\Omega^{-1/2}~.
\end{equation}
$\Delta\ell\sim 300 ~\Omega^{-1/2}$ favors an open inflationary model,
whereas $\Delta\ell\sim 300$  favors a topological defects model with
$\Omega=1$. 

\subsection{Results from specific models}

Studying the acoustic peaks for perturbations induced by global textures 
and CDM, we have found~\cite{ram} that the height of the first acoustic 
peak is smaller than in standard CDM models, and that its position is 
shifted to smaller angular scales. More precisely, we found~\cite{ram} that
the amplitude of the first acoustic peak is only $\sim 2.5$
times higher than the SW plateau and its position is at $\ell\sim 360$.
We believe that our results are basically valid for all global topological
defects.
On the other hand, standard inflation predicts the position of the first
peak at $\ell\sim 220$ and its amplitude $\sim(4-6)$ times higher than 
the SW plateau~\cite{stein}. There are however non-generic, open, tilted
inflationary models which might reproduce similar signature in the CMBR angular
power spectrum, as global topological defects models.

Some studies have been also done in the case that density fluctuations
are seeded by local cosmic strings~\cite{chm}. It was found that the matter
power spectrum is very sensitive to the assumptions made about string decay
(e.g., gravitational radiation; very high energy particles).
Numerical simulations showed that the peak of the spectrum is always at 
smaller scales that standard CDM predictions, or observations.
At scales 100 Mpc/h, which are unaffected by non-linear gravitational
evolution, the bias factor is unreasonably high (its value depends on string
decay products); but this is a less severe problem than for global defects.
The addition of a cosmological constant leads to a better agreement
with data for the cosmic string model of large scale structure 
formation~\cite{cs2}.
The CMBR power spectrum is relatively insensitive to the equation
of state of the extra fluid. Numerical simulations~\cite{chm} revealed
a reasonably high single acoustic peak at $\ell=400-600$, following a 
pronouncedly tilted large angle plateau.

Numerical simulations of local cosmic strings~\cite{abr,cs} 
and global ${\cal O}(N)$ defects~\cite{ram,pst,fit}
led to the conclusion that these models cannot reproduce
the characteristics of the acoustic peaks,
as they have been revealed by currently available data. Of course one
has to keep in mind that the experimental data have error bars which are still 
rather considerable. Some authors already
concluded~\cite{abr,pst} that models where structure formation is
triggered by scaling causal defects are ruled out.

However, it may still be too early to rush into conclusions.
In a simple generic parameterization of the energy momentum tensor
of two families of more general scaling causal seeds models~\cite{fit},
we were able to fit very well the available data. More precisely, 
we computed~\cite{fit} CMBR angular power spectra for two scaling causal
seeds models inspired by global topological defects: 
${\cal O}(4)$ texture models and the large-$N$ limit of ${\cal O}(N)$ 
models. Our aim was to investigate whether the available measurements
of CMBR anisotropies could rule out a generic class of seed perturbations
constrained just by energy momentum conservation and scaling arguments.
Using ${\cal \chi}^2$ fitting to compare the predictions of our models to 
flat-band power measurements of the CMBR, we found~\cite{fit} that the 
answer to this question is {\sl no}.
Causal scaling seeds perturbations have been also analyzed in Ref.~[26],
ignoring for the time being the physical origin of these seeds,
something which people are almost always doing in the context of 
inflationary perturbations. The results of Ref.~[26] agree with our
conclusions~\cite{fit}.

The most severe
problem for topological defects models of structure formation is their
predicted~\cite{abr,pst} lack of large scale power in the matter power
spectrum, once normalized to COBE. 
 Choosing scales of $100$ Mpc/h, which are most probably
unaffected by non-linear gravitational evolution, standard topological
defects models, once normalized to COBE, require a bias factor
($b_{100}$) on scales of $100$ Mpc/h of $b_{100}\approx 5$, to
reconcile the predictions for the density field fluctuations with the
observed galaxy distribution. However, the latest theoretical and
experimental studies favor a current value of $b_{100}$ close
to unity.

The global texture model of structure formation in cosmological
models with non-zero cosmological constant and different values of 
the Hubble parameters has been investigated in Ref.~[27]. 
The authors deduced that the absence of significant 
acoustic peaks in the CMBR anisotropy spectrum is a robust result
for all models with global textures, as well as the large-$N$ limit 
of ${\cal O}(N)$ models, for all considered choices of cosmological 
parameters~\cite{mra}. 
More precisely, on intermediate angular scales these models
are unable to fit the measurements of the Saskatoon experiment. 
Moreover, the dark matter power spectrum on large scales
($\gsim 20$ Mpc/h) is considerably lower than the measured galaxy
power spectrum~\cite{mra}. However, here it still remains open 
the question of the 
biasing problem. It seems that the rejection of models with global textures, 
as well as the large-$N$ limit of ${\cal O}(N)$ models comes once
we consider the bulk velocity on large scales. More precisely,
the large scale bulk velocities are by a factor of about 3 to 5 
smaller 
than the value inferred from the peculiar velocity data~\cite{mra}.

At this point, I would like to draw your attention that, as it has been 
emphasized in Ref.~[28], all presently available
numerical simulations of topological defects models have overlooked
the defect decay into gravitational radiation and/or elementary
particles. As it was shown in Ref.~[28], the predictions regarding 
the degree-scale amplitude of the CMBR anisotropies, and the shape
of the matter power spectrum change dramatically, once one considers
that a fraction of the energy of a network of topological defects 
is released directly into photons, baryons and neutrinos.
This conclusion seems to be independent of the particular
type of topological defects and/or their decay process.

\subsection{Lessons we have already learned  and lessons
we expect to learn from future experiments of the cosmic 
microwave background  anisotropies measurements }

The present results from the CMBR measurements lead to the following 
conclusions:\\
1. The high level of isotropy of the CMBR favors the FLRW metric on 
large scales.\\
2. The matter component of the universe is consistent mostly of cold dark 
matter, with baryons, perhaps massive neutrinos, plus curvature and/or vacuum 
components.\\
3. The COBE data give the normalization of models of the large scale 
structure. The large scale anisotropies measure the amplitude of the Bardeen 
potentials on large scales, which provides the normalization of the matter 
power spectrum. It results that the Bardeen potentials have to be dominated by
dark matter, which is not coupled to photons  which prevent baryons from
collapse before recombination. The galaxy distribution in the local universe
shows that matter is formed ``bottom up'', implying that the  velocity
dispersion of dark matter must be extremely small, and therefore, it must
be mostly cold.\\
4. There are fluctuations at $\ell \gsim 100$, implying that there is no 
early ionization.\\
5. The CMBR anisotropy is linearly polarized at a very small level.\\
6. Scalar (S), vector (V), tensor (T) fluctuations would be produced at early 
times in roughly equal amounts. Vector modes decay with time. Today there are
tensor and scalar modes with $T/S <1$. If tensor modes have an almost
scale-invariant spectrum, then the possibility to detect gravity waves
with LIGO/LISA is small.

The future experiments will allow us:\\
1. To determine to a precision of one percent or better the cosmological parameters.\\
2. To measure the polarization over a range of scales.\\
3. To learn about early universe physics.\\
4. To learn about non-linear astrophysics.

\section{Conclusions}

On large angular scales, both, topological defects models as well as inflation
lead to approximately scale invariant spectra. Thus, we cannot discriminate 
among them using the measurements of the COBE satellite. We need to probe 
anisotropies on smaller angular scales. The simplest topological defects
models appear to have some conflicts with observational data. 
More precisely,
they cannot reproduce the data of the Saskatoon experiment and they predict 
lack of large scale power in the matter power spectrum, once normalized to 
COBE (they require a high bias factor). 
However, taking into account some micro-physics regarding the decay of 
topological defects, in particular in the case of superconducting
cosmic strings, the predictions regarding the degree-scale amplitude
of CMBR anisotropies as well as the shape of the matter power spectrum,
can be considerably modified.

If acoustic peaks are confirmed and if the 
first peak is  at $\ell < 300$ with an amplitude $\sim (4-6)$ times higher 
than the SW plateau, then the simplest topological defects
 models are ruled out. If non-Gaussian statistics are 
confirmed and if topological defects models as the mechanism to induce the
initial perturbations are ruled out, then one should take seriously
non-vacuum initial states for cosmological perturbations of quantum mechanical
 origin. If the power spectrum looks different than what standard inflation 
predicts, then one should investigate more complicated models, like
for example, double or multiple~\cite{sarkar,mn}
inflation models, models with a kink in the inflaton potential, broken scale
invariant models~\cite{polarski}, and/or models where the initial state 
has a built-in characteristic scale~\cite{jam}.

I believe that it is still early to rush into definite conclusions
regarding the origin of the initial perturbations. The simple models
we have in mind from each one of the two families of models (inflation
or topological defects) run into problems once compared with the
currently available 
data. Unfortunately, due to their non-linearity, defects models are
more difficult to handle and therefore they have been less investigated.
 Further observational and experimental data, as well as
more profound theoretical studies, are necessary before we know
for sure what it is still hidden behind the origin of the large scale 
structure in the universe.

\vspace*{-2pt}
\section*{Acknowledgments}
It is a pleasure to thank the organizers of the 1999 Spanish Relativity 
Meeting at Bilbao, 
for inviting me to give this talk, as well as for the friendly and
pleasant atmosphere during the meeting.

\end{document}